\documentclass[sigconf]{acmart}

\usepackage{booktabs} 
\usepackage{paralist}
\usepackage{subfig}
\usepackage{xspace}
\usepackage[normalem]{ulem}

\newcommand{\DAGs}{\textbf{DATs}\xspace}

\newcommand{\organization}{organization\xspace} 
\newcommand{\detr}{detailed access trajectory\xspace}
\newcommand{\DET}{DAT\xspace}
\newcommand{\dets}{detailed access trajectories\xspace}
\newcommand{\DETS}{DATs\xspace}
\newcommand{\reviewAfterALongTime}{``\textit{return after long time}''\xspace}
\newcommand{\returnToPreviouslyWatched}{``\textit{return to most recently watched}''\xspace}
\newcommand{\returnToPreviouslySkipped}{``\textit{return to previously skipped}''\xspace}

\newcommand\defineshortcoursenames[2]{%
  \expandafter\newcommand\csname var#1var\endcsname{#2}%
}
\newcommand{\shortcoursename}[1]{\csname var#1var\endcsname}
\defineshortcoursenames{1A}{\textit{1A}}
\defineshortcoursenames{1B}{\textit{1B}}
\defineshortcoursenames{1C}{\textit{1C}}
\defineshortcoursenames{2A}{\textit{2A}}
\defineshortcoursenames{2B}{\textit{2B}}
\defineshortcoursenames{2C}{\textit{2C}}

\setcopyright{acmcopyright}

\acmDOI{}

\acmISBN{}

\acmConference[LAK'19]{International Conference on Learning Analytics \& Knowledge}{March 2019}{Arizona, US}
\acmYear{2019}
\copyrightyear{2019}

\begin{document}
\title{Using Detailed Access Trajectories for Learning Behavior Analysis}

\author{Yanbang Wang}
\affiliation{%
  \department{Department of Computer Science and Engineering}
  \institution{The Hong Kong University of Science and Technology}
  \streetaddress{Clear Water Bay, Kowloon}
  \city{Hong Kong SAR}
  \country{China}
}
\email{ywangdr@connect.ust.hk}

\author{Nancy Law}
\affiliation{%
  \department{Faculty of Education}
  \institution{The University of Hong Kong}
  \streetaddress{Pokfulam}
  \city{Hong Kong SAR}
  \country{China}
}
\email{nlaw@hku.hk}

\author{Erik Hemberg}
\affiliation{%
  \department{Computer Science and Artificial Intelligence Laboratory}
  \institution{Massachusetts Institute of Technology}
  \streetaddress{32 Vassar Street}
  \city{Cambridge}
  \state{MA}
  \country{USA}
}
\email{hembergerik@csail.mit.edu}

\author{Una-May O'Reilly}
\affiliation{%
  \department{Computer Science and Artificial Intelligence Laboratory}
  \institution{Massachusetts Institute of Technology}
  \streetaddress{32 Vassar Street}
  \city{Cambridge}
  \state{MA}
  \country{USA}
}
\email{unamay@csail.mit.edu}

\begin{abstract}
Student learning activity in MOOCs can be viewed from multiple perspectives. We present a new \organization of MOOC learner activity data at a resolution that is in between the fine granularity of the clickstream and coarse organizations that count activities, aggregate students or use long duration time units. A \textit{detailed access trajectory} (\DET) consists of binary values and is two dimensional with one axis that is a time series, e.g. days and the other that is a chronologically ordered list of a MOOC component type's instances, e.g. videos in instructional order. Most popular MOOC platforms generate data that can be organized as \dets (\DETS).  We explore the value of \DETS by conducting four empirical mini-studies. Our studies suggest \DETS contain rich information about students' learning behaviors and facilitate MOOC learning analyses.
\end{abstract}

%
%



\maketitle

\section{Introduction}
The analysis of students' learning behavior has been a major focus for MOOC learning analytics\cite{zhuoxuan2015learning, rai2016influencing, Davis:2018:HMK:3170358.3170383, boroujeni2018discovery}. Popular MOOC platforms, like Edx and Coursera, usually provide comprehensive click-stream logs of all interactions with the MOOC platform or organized data in BigQuery tables. This data enables us to perform learning behavior analytics at many different granularities and behavior categories. Aggregation is often an efficient approach to the analysis of the large quantity of students and activity data.  Existing works have examined a wide range of perspectives, including general group behaviors and detailed event-wise individual browser post and get requests. 

\begin{figure}
\centering
\includegraphics[width=0.49\textwidth]{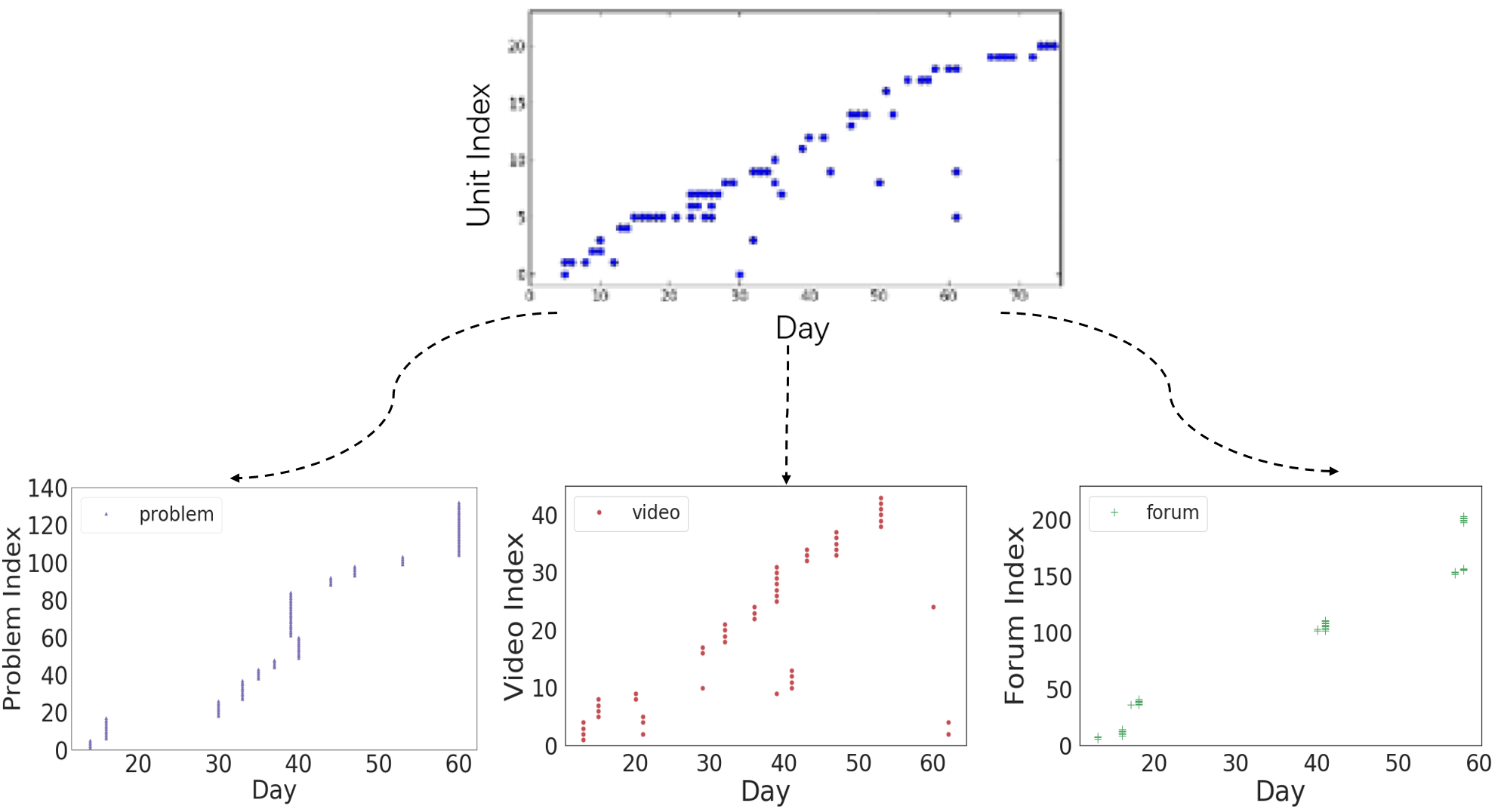}
\caption{Aggregated Activity Plot from \cite{halawa2014dropout}(top) and \DETS (bottom) for problem submission, video watching and forum participation. All plots show the number of days from the start of the course on the horizontal axis and the component accessed on the vertical axis.}
\label{fig:basic_repr}
\end{figure}

Here, we study MOOC learning behaviors with a \textit{\detr} (\DET). This representation allows us to study learning patterns and behaviors from the perspective of \textit{when} a \textit{particular student accesses a particular component} assuming the component is in an ordering by when it appears in course material. The \DET representation is inspired by \cite{halawa2014dropout} where student learning trajectories are visualized as a step-like signal. This is shown to the top plot in Figure ~\ref{fig:basic_repr} as an ``activity plot''. It shows the number of days from the start of the course on the horizontal axis and the unit access on the vertical axis. E.g. if on a day $d$, a student accesses material from unit $u$, then the plot has a mark at coordinate $(d, u)$. This reflects how a student proceeds through the course, accessing unit material over time. \DETS expand the original work by detailing one type of the course component, e.g.:\begin{inparaenum}[\itshape A)]
\item video watching,
\item problem submission, or
\item active and/or passive forum participation (reading and submitting).
\end{inparaenum}
The \DET gives insights on how many and different course components are viewed, skipped and revisited, when this happens, the length of time that a student is absent, when a student stops out, etc.  In this paper, we leverage the advantages of \DET for three explorations:
\begin{asparaenum}
\item \textbf{Learning Behavior Patterns} We visualize video watching \DETS and observe a distinctive behavior pattern where the last video of the previous day is the first video the student watches on the next day. We ask whether this behavior, which could be interpreted as either  knowledge reinforcement or video watching completion, is correlated to grade. We also observe two more distinctive patterns where a video introducing material early in the course is revisited much later on, or is skipped for the first time but revisited much later on. These behaviors, again, could have multiple interpretations. We investigate their correlation to grade also.\label{question:LBP}  
\item \textbf{Learning Design Patterns} We probe the possibility that \DETS can inform learning design hypotheses about \textbf{L}earning \textbf{D}esign \textbf{P}atterns.  This is a core concern of designers and instructors when they are designing their course for efficient student learning.\label{question:LDP}
\item \textbf{Background Examination} We use \DETS to examine students from educational and geographical backgrounds that make them potentially marginalized. The \DET helps identify whether these students are struggling with their MOOC studies, potentially allowing them to receive appropriate help.\label{question:Margin}  
\item \textbf{Dimensionality Reduction} We ask how a large quantity of \DETS can be summarized and mapped into a low dimensional embedding that allows them to be input to modeling or be analyzed with 2D visualization yielding potential for observing clusters\label{question:Embed} 
\end{asparaenum}

The rest of the paper is structured as follows: Section~\ref{sec:related-work} presents related work. The courses we use for demonstration are introduced in Section~\ref{sec:dataset}. Section~\ref{sec:detDefn} defines a \DET. Section ~\ref{sec:combinatorial} covers Exploration(\ref{question:LBP}). Section ~\ref{sec:inclusion} covers Exploration(\ref{question:Margin}). Section~\ref{sec:vdag} we present findings to Exploration(\ref{question:Embed}). Finally, Section~\ref{sec:conclusions--future} concludes and discusses future work.

\section{Related Work}
\label{sec:related-work}

\subsubsection*{Explorations(\ref{question:LBP}, \ref{question:LDP}) Learning Patterns}
There has been a lot of research over the past two decades on teaching as a design science~\cite{laurillard2013teaching}. Some work in studying learning design in the context of learning technologies has been inspired by~\cite{alexander1979timeless}'s concept of design patterns, which are ``invariants'' underlying successful designs. In the context of learning design, the core elements of a design pattern comprise descriptions of the ``problem'' (the learning outcome to be achieved), the context (the learning situation, including the course and student contexts), and the ``solution'' (the sequences of learning activities involving tangible, virtual and social interactions). Learning design patterns (LDP) are the (not necessarily conscious) assumptions teachers have about how students should interact with specific materials and engage in designated learning activities for effective learning~\cite{law2017pattern}.  Learning behavior patterns (LBP) are, in contrast, what students actually do.  By identifying empirically the localized learning behavior patterns exhibited by students, teachers can find out:
\begin{inparaenum}[\itshape (1)]
\item what proportion of the students actually exhibit behavior as intended by the teacher's LDP, and whether those following the intended LDP exhibit better learning outcomes;\item what other learning behavior patterns exist, and whether any of these patterns are strongly correlated with students' learning success or failure;
\item whether students' adoption of the observed patterns were dependent on their contextual backgrounds, and
\item whether the effectiveness of the observed learning behavior patterns interact with students' contextual background.
\end{inparaenum}
Answers to the above questions would make significant contributions to teachers' learning design knowledge and practices, providing evidence-based input to personalizing learning design that are sensitive to both the specific learning objectives targeted and the students' contexts. 

\subsubsection*{Exploration(\ref{question:Margin}) Background Examination}\label{subsec:bg_margin}

Previous works have noted that students on popular MOOC platforms have highly diverse backgrounds~\cite{deboer2013diversity,pursel2016understanding,deboer2013bringing}. Some studies took one step further by examining the correlation between students' background and their learning behaviors. For example, \cite{HOOD201583} analyzed survey day from MOOC users and found that those with strong data science backgrounds differ significantly with other students in terms of their self-regulated learning;\cite{guo2014demographic} investigated how students' demographic background could affect their navigation strategies, and found that older students and students from countries with low student-teacher rate are more likely to do follow a steady learning pattern.

Though the connection between background and learning behaviors are widely studied, very few works studied marginalized group's learning behaviors by analyzing their learning data. Many studies have mentioned the importance of studying marginalized student groups \cite{brugha2016examining,wilson2018online,mcandrew2013open}. 

\subsubsection*{Exploration(\ref{question:Embed}) Dimensionality Reduction}
\label{sec:repr_learn}
The compact representation of a \DETS is variable length (and the matrix representation impractical).
Finding a fixed length, numerical vector that could represent a \DET would support its use in existing modeling contexts, such as \begin{inparaenum}[\itshape 1)]
\item  predicting grades by student's learning behaviors \cite{ren2016predict,7452320,7313031,Xu:2016:MCG:2934357.2934361};
\item student grouping (clustering) or subpopulation analysis \cite{Corrin:2017:ULA:3027385.3027448,Kizilcec:2013:DDA:2460296.2460330,Ferguson:2015:EEA:2723576.2723606};
\item transfer knowledge about student populations across courses\cite{10.1007/978-3-319-19773-9_6,He:2015:IAS:2886521.2886563}.
\end{inparaenum}
These works require a numerical vector representations as input and reply upon by calculating a number of learning-related features (e.g. number of watched videos, frequency of login, etc.). Three major problems exist with such method: 
\begin{asparaenum}
\item Many features are highly correlated with each other. For example, "number of watched videos" is a ubiquitously used strong feature that is highly correlated with other features such as "number of assignment submissions", "number of forum posts", "number of video pauses", etc.(used by\cite{Corrin:2017:ULA:3027385.3027448})
\item Since all the features are manually designed, very often some aspects of learning behavior are subjectively overemphasized, while some others are ignored. This problem along with 1 often leads to strong bias in the final feature vector.
\item The handcrafted features are usually high-level statistical aggregations. However,  a lot of information is contained in shorter time windows, such as the periodicity of material access and frequent material revisits over a short time. The manually designed features usually fail to capture such subtleties. 
\end{asparaenum}

\section{Demonstration Courses}
\label{sec:dataset}
We analyze two courses on Edx
\begin{inparaenum}[\itshape A)]
\item MITx 6.00.1x \textit{Introduction to Computer Science and Programming Using Python},
\item MITx 6.00.2x \textit{Introduction to Computational Thinking and Data Science}.
\end{inparaenum}
Each course has three offerings in 2016 and 2017. The student population and total activity of each offering varies, see Table~\ref{table:courseStatistics}, with diverse demographics. Each offering lasts 10 weeks. The final grade is the weighted sum of scores in finger exercises (weight = 0.1), problem sets (weight = 0.4), midterm or quiz (weight = 0.25), and final exam (weight = 0.25).  Both courses have multiple units, where each unit has an associated graded problem set. Students are expected to watch lecture videos narrated by instructors and complete``finger exercises'' - optional problems interspersed in lecture videos that teach the content discussed in the video.  Forum participation is optional and each discussion forum contains thousands of posts. The topics of each course differ because one course is the continuation of the other. The quantities of videos and finger exercises is much higher in 6.00.1x, see Table~\ref{table:numberofitems}. 
\begin{table}[t]
\begin{center}
  \caption{Number of students and log events for 6.00.1x and 6.00.2x. We use the notation in brackets as identifiers throughout the paper.}
\begin{tabular}{lcc}
 \toprule
 Course  & \#Students & \# Log Events \\
 \midrule
6.00.1x Summer 2016A(\shortcoursename{1A}) & 113,099 & 17,333,974 \\
6.00.1x Summer 2016B(\shortcoursename{1B}) &  40,727& 7,900,908\\
6.00.1x Spring 2017(\shortcoursename{1C}) & 69,399   & 13,176,220 \\
6.00.2x Spring 2016(\shortcoursename{2A}) & 18,362 & 2,642,528  \\
6.00.2x Fall 2016(\shortcoursename{2B}) & 22,023 & 2,501,276 \\
6.00.2x Spring 2017(\shortcoursename{2C}) & 18,281 & 2,034,539 \\

 \bottomrule
\end{tabular}
\label{table:courseStatistics}
\end{center}
\end{table}

\begin{table}[t]
\begin{center}
  \caption{Resource quantities in terms of video and finger exercises, with 6.00.1x having many more than 6.00.2x.}
\begin{tabular}{lcc}
 \toprule
 Course  & \# Videos & \# Problems \\
 \midrule
6.00.1x & 81 & 212 \\
6.00.2x & 43 & 156 \\

 \bottomrule
\end{tabular}
\label{table:numberofitems}
\end{center}
\end{table}

For data preprocessing, for each offering, we reference multiple \texttt{BigQuery} tables and extract three \DETS for each student using tables named, respectively:
\begin{inparaenum}[\itshape A)]
\item \texttt{video\_stats\_day},
\item \texttt{person\_problem},
\item \texttt{forum\_person}.
\end{inparaenum}
These are equivalent to video accesses, problem set accesses and forum accesses (read or write). The generation of the first two \DETS is straightforwardly  done on Coursera platform but a forum-participation \DET is slightly more platform-dependent.

\section{Definition: Detailed Access Trajectory}\label{sec:detDefn}
A \DET is logically envisioned as a 2D matrix $(DxN)$ where the row dimension is course days $(1, \dots, D)$) and the columns are ordered course components from $(1, \dots, N)$. The ordering is how the components are presented within the course structure.  At $(d_i,c_j)$ is either a $1$ or $0$ to denote the student accessing the j'th component on the ith day.  Because this matrix would be sparse and large, we use a compact representation that is a series that expresses only the entries set to 1 (for access) by their day and component indices. 

\section{Learning Patterns}
\label{sec:combinatorial}

\subsection{Learning Behavior Patterns (LBP)}
\label{subsec:localpattern}
\begin{figure}[tb]
\includegraphics[width=0.32\textwidth]{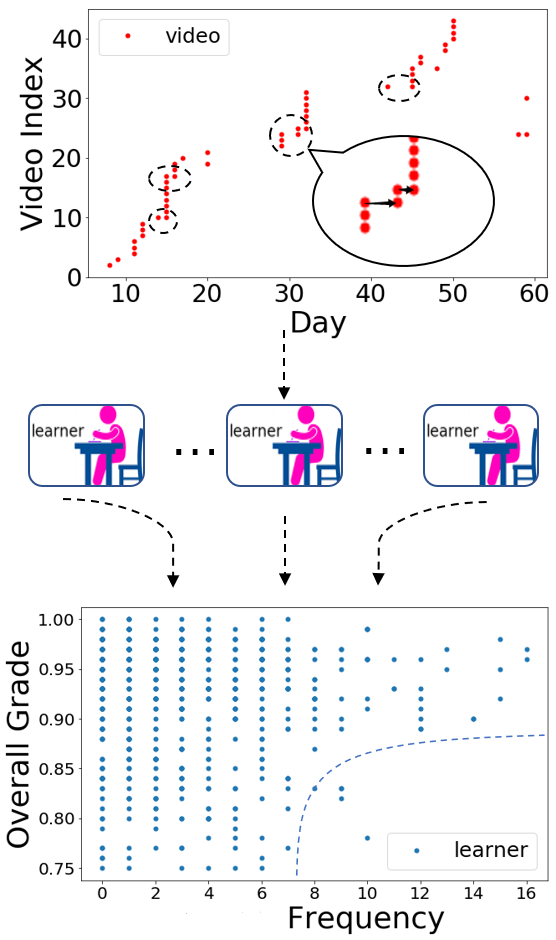}
\caption{LBP: \returnToPreviouslyWatched (Course \shortcoursename{2C}). Top: \DET of a student "recapping" earlier videos before final exam; Bottom: Scatter plot of grade(y-axis) and number of recaps(x-axis). Students frequently "recapping" tend to have higher grades.}
\label{fig:recap}
\end{figure}

The point of this exploration is to determine if any learning behavior pattern (LBP) we can discern by visualizing the DAT is informative to teachers who start from assuming a LDP.
We visualize video watching \DETS to look for multiple "localized video watching patterns". We observe a distinctive behavior pattern where the last video of the previous day is the first video the student watches on the next day.  For simplicity, we dub it \returnToPreviouslyWatched.  When visualized the pattern looks like a step signal, as shown by the top plot of figure \ref{fig:recap}. It shows that the student seems to review or complete learning previously accessed knowledge when he/she starts learning everyday. We ask whether this behavior, which could be interpreted as either  knowledge reinforcement or video watching completion, is correlated to grade. 

\begin{figure}[tb]
\includegraphics[width=0.32\textwidth]{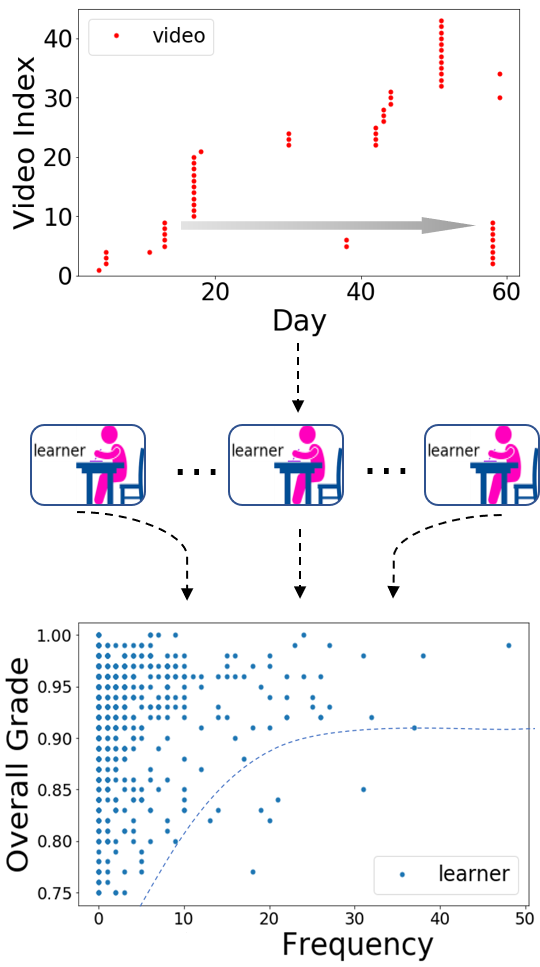}
\caption{LBP: \reviewAfterALongTime (Course \shortcoursename{2C}). Top: \DET of a student replaying earlier videos before final exam; Bottom: scatter plot of grade (y-axis) and number of replays(x-axis). Students with frequent "replaying" behavior tend to have higher grades.}
\label{fig:local_video_lbp_2}
\end{figure}

Adopting the same method, we also observe two more distinctive patterns in this way 1)\reviewAfterALongTime:  a video introducing material early in the course is revisited much later on; 2) \returnToPreviouslySkipped: a video introducing material early in the course is skipped over at the first but is revisited much later on. Both behaviors, again, could have multiple interpretations. For \reviewAfterALongTime one possible interpretation is that the student is actively learning, explicitly deciding to review previous material. Another is that the video is left unfinished. To distinguish this behavior with the \returnToPreviouslyWatched behavior, we only consider a pattern as \reviewAfterALongTime when a person re-watches a video after at least one active day when he/she did not watch that video, illustrated by the top plot of figure \ref{fig:local_video_lbp_2}. \returnToPreviouslySkipped possibly illustrates a latent pattern of active learning: a student realizing that certain lecture videos skipped early on for some reasons actually matter, and so he/she consciously seeks those lecture videos to pick up the missing knowledge. 

\begin{figure}[tb]
\includegraphics[width=0.35\textwidth]{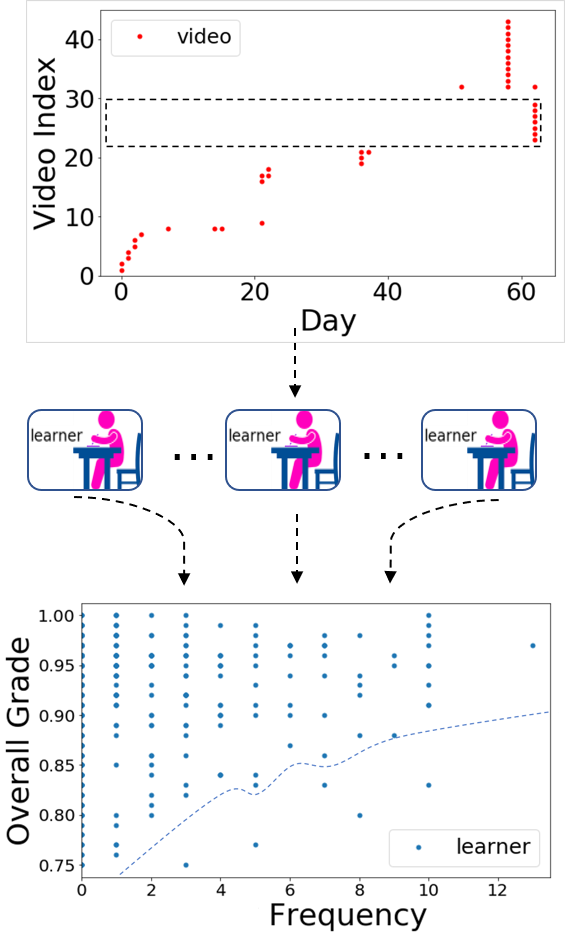}
\caption{LBP: \returnToPreviouslySkipped (Course \shortcoursename{2C}). Top: \DET of a student making up earlier skipped videos before final exam; Bottom: scatter plot of grade (y-axis) and number of replays(x-axis). Student with frequent behavior tend to have higher grades.}
\label{fig:local_video_lbp_3}
\end{figure}

Scatter plots on the bottom of the three figures visualize the correlation between occurrence frequencies of the local pattern and grade. Figure \ref{fig:recap} shows that most students do not \returnToPreviouslyWatched video very often. While no significant difference in grade distribution can be observed on students that do it occasionally, students who \returnToPreviouslyWatched more regularly often get high grades. This intriguing pattern is observed on all the six offerings, which suggests that when a student's \returnToPreviouslyWatched learning behavior is regular, that student has a strong correlation with high course grade. Figure \ref{fig:local_video_lbp_2} and \ref{fig:local_video_lbp_3} can also be interpreted in a similar way (observe that grades of student with high learning pattern frequency in Figure \ref{fig:local_video_lbp_2} and \ref{fig:local_video_lbp_3} concentrate on the upper half of the plot respectively). This indicates that a student has a higher likelihood to receive a high grade in the course with more frequent \reviewAfterALongTime and \returnToPreviouslySkipped behavior. 

Further statistical testing is conducted to verify our observations as well as to determine the exact cutoff pattern frequencies that most prominently affect students' grade. This is done by iterating though all possible values of cutoff frequencies (from 1 to \textit{max pattern frequency} $-$ 1) and selecting one that results highest p-values of one-tailed t-test. Table \ref{table:stats_lbp} summarize the aforementioned cutoff frequencies of LDP that affect grade, and their corresponding p-values that measure how strongly students' grades are affected by exhibiting learning pattern frequencies below or above the cutoff value. 
\begin{table}[t]
  \footnotesize
\begin{center}
  \caption{Cutoff frequencies that affect grade distribution and corresponding p-values that are all statistically significant ($<$0.05). N=591. }
\begin{tabular}{lcc}
 \toprule
LBP  & Cutoff Frequency &p-value \\
 \midrule
\returnToPreviouslyWatched  & 7 & 0.014\\
\reviewAfterALongTime & 10 & 0.033 \\
\returnToPreviouslySkipped & 8 & 0.012 \\

 \bottomrule
\end{tabular}
\label{table:stats_lbp}
\end{center}
\end{table}

\subsection{Learning Design Pattern}

Learning is not only about how much effort one spends on learning,
but also about how one distributes the effort. Various hypotheses
and theories exist regarding specific patterns of learning materials access and learning activity engagement that
student can follow to achieve better course performance.
Course instructors can design their courses based on pedagogical theories and their own professional experience to encourage such learning patterns and thus help students learn
better. Design patterns that instructors adopt in their course design for students to follow we refer to as 
\textbf{learning design pattern (LDP)}~\cite{law2017pattern}\label{ldp}.

LDPs are usually difficult to evaluate and justify with traditional
experimental methods, due to limited observation size and data
collection difficulties. The advent of MOOCs provides valuable opportunities
to examine LDPs on a larger scale with more visual
and statistical analysis. For example, education experts
and the MITx course designer have jointly identified a LDP regarding
watching video and participating in forum discussion in a
the MITx 6.00.1x course. Course instructors from MIT
identified similar LDPs in the other computational-thinking course we investigate. Supported by \DETS, we perform an exploration of one LDP identified by experts. The LDP states that "students should participate in forum discussions shortly after they watch a video". Our analysis will help answer
the following questions:
\begin{asparaitem}
\item How many student exhibit the learning behavior pattern that aligns with the LDP? Is the
number significant enough relative to the course population?
\item What are the grades of student who did or did not exhibit such LDP aligned behavior patterns?
\item Are there any ambiguities in the definition of the LDP? How might such ambiguities affect designer conclusions?
\end{asparaitem}

\begin{figure}[t]
\includegraphics[width=0.49\textwidth]{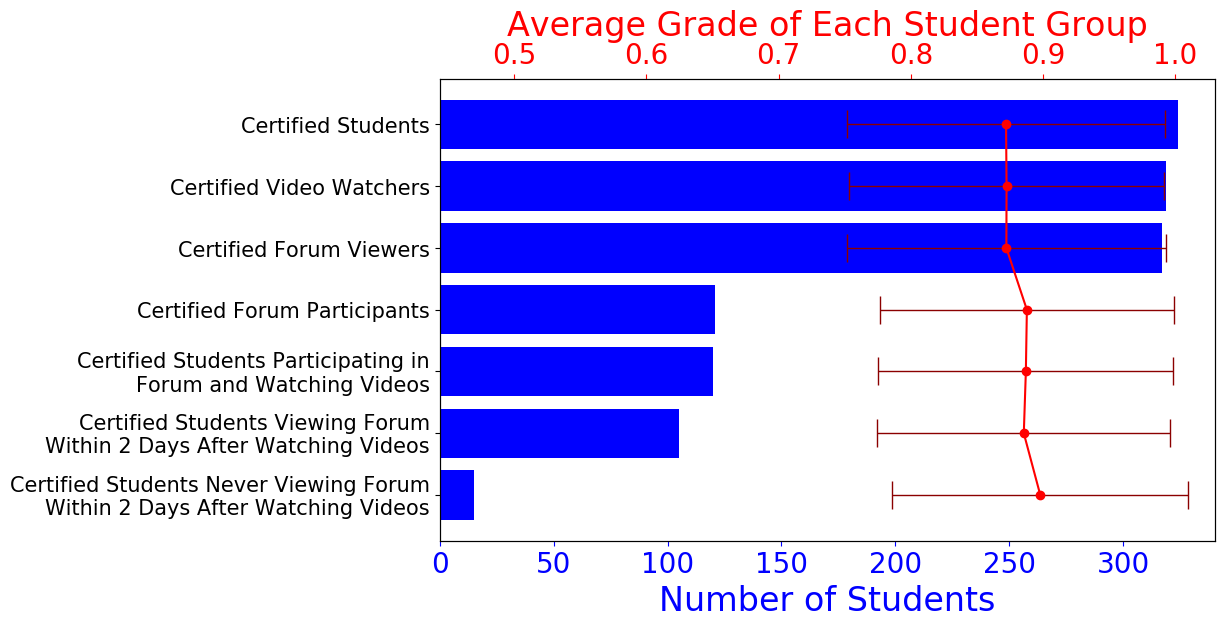}
\caption{Number of different groups of participants and their corresponding average grades (Course \shortcoursename{2C}). Blue bars and bottom horizontal axis shows the number of students in the group, and red components with top horizontal axis mark the average grades of the group and the variance.}
\label{fig:ldp}
\end{figure}

We start by counting the number of different groups of certified participants\footnote{We restrict our exploration to certified student, who received a certificate for finishing the course after the course end.} including video watchers, forum viewers and participants, etc. We calculate the average overall course grades (on a unit scale) of student within each group, along with their grade variance. The aggregations are visualized in figure \ref{fig:ldp}. We can see that in the smallest offering, MITx 6.002 spring 2017, almost all certified course participants watched the videos and viewed the forums. However, less than half of the certified student participate in the forum discussions. 

With regard to the specific learning design pattern, we see from the fifth and sixth bar that most student that watch videos and participate in forums follow the specific learning design pattern\footnote{There is ambiguity in the statement as to what exactly "shortly" means: it could refer to a time lag of one, two or three days, or even longer.}.

The red line in figure \ref{fig:ldp} indicates that average grades do not vary significantly among different groups of student, especially on the last two rows that we care most about. We then performed two-tailed t-test on the last two groups:
\begin{asparaitem}
\item $H_0$: \textit{Students viewing forums within 2 days after watching videos (Group Y, hereafter) have the same average as the students never viewing forums within two days after watching videos (Group N, hereafter)} ;
\item  $H_1$: \textit{Group Y does not have the same average grade as Group N.} 
\end{asparaitem}
The p-value is 0.6841, which is statistically insignificant. The same analysis was repeated on the other 5 offerings with no significant differences. This leads us to conclude that given the large variance of within-group grade (as visualized by the red error bars), no strong correlation is observed between the LDP and grade. Previous studies similarly mentioned about weak correlation between forum participation and grade among students who pass the course \cite{Wise:2018:URD:3170358.3170403}.

Despite the insignificance, we can still investigate this LDP in more detail. We ask the question: are students more likely to go to a forum shortly after watching a lecture video? Notice that the key phrase "shortly" is imprecisely defined. In previous analysis, we heuristically set the length of "shortly" to be two days.  Here, "shortly after" is further parameterized to an offset of n days, where n takes an integer value\footnote{$n$ can be negative --- in that case the LDP becomes "viewing a forum $|n|$ days before watching a lecture video"}. 
We define two events for a student \footnote{1) The problem is modeled such that for a fixed $n$, every student the identical independent distribution of the following events. 2) We discussion only students that both watch videos and view forums}:
\begin{asparaitem}
\item Event $A$: the student watches at least one video on day $x$;
\item Event $B$: the student views at least one forum thread on day $x+n$.
\end{asparaitem}
$x$ could be any integer number within the range of course duration\footnote{For simplicity, we ignore the edge effect of $x$ approaching the beginning or end of the course}. Bernoulli random variable $A, B$ respectively marks the probability that event $A, B$ happens (r.v. $A$ and event $A$ used interchangeably hereafter, same for $B$). An unbiased estimation of $P(B|A)$ for fixed n can therefore be obtained from a \DET with 3 steps:
\begin{asparaenum}
\item For each student, count the number of video-watching days $v\_days$ from the student's video-watching \DET. For each video watching day, check forum \DET to see if there is a forum view record on n-th day after that day. Count the number of video watching days that have such a paired forum-view day, $v\_f\_days$.
\item Sum up $v\_days$ and $v\_f\_days$ for all students to obtain  $total\_v\_days$ and $total\_v\_f\_days$;
\item $\hat{P}(B|A)$ =  $total\_v\_days / total\_v\_f\_days$;
\end{asparaenum}

\begin{figure}[t]
\includegraphics[width=0.45\textwidth]{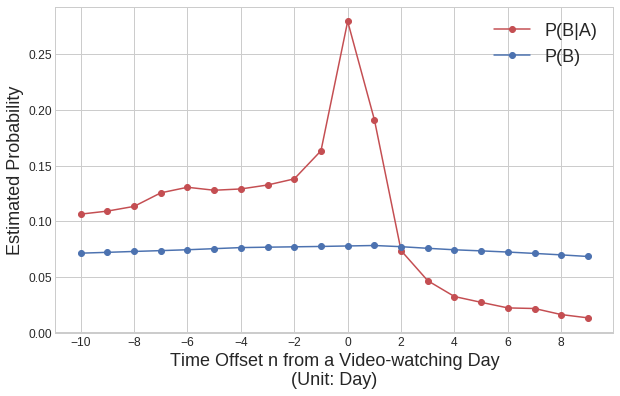}
\caption{Estimated conditional probability that a day with time offset n is a forum-viewing day given that the day with offset 0 is a video-watching day.}
\label{fig:forum_prob_offset}
\end{figure}

With this analysis we find that $N$ for the smallest offering Course 2C is 20,541, which is statistically significant with a small 99\% confidence interval. Figure \ref{fig:forum_prob_offset} plots the estimated conditional probability $\hat{P}(A|B)$ (red) against corresponding time offset parameter $n$ in Figure \ref{fig:forum_prob_offset}. It is clearly observed that $\hat{P}(B|A)$ peaks at zero offset, and then drops rapidly sideway, meaning that if  a random student in the course watches a video, then the student has the highest probability of viewing a forum on the same day, compared to other days within a range of 10 days centered around. The second highest probability is exactly one day after watch a video. Similar patterns are observed in all other offerings.
 
The blue line in figure \ref{fig:forum_prob_offset} further shows the estimated natural distribution of r.v. $B$ with $N=153,792$. It means that on a random day, a random student has  around 0.075 probability to view at least one forum. More importantly, it could be clearly observed that the $\hat{P}(B|A)$ and $\hat{P}(B)$ has very different distribution for each offset n. Further KS tests show that we have p-value < 0.0001 to reject this hypothesis. Therefore, with high confidence level we conclude that $P(B|A) \neq P(B)$, and thus $A$ and $B$ are dependent. In other words, we conclude that the LDP is a strong pattern, though its grade implication is quite weak.

\section{Under-Represented Student Groups}
\label{sec:inclusion}

Few works have concretely identified different potentially marginalized groups. In this section we identify several under represented groups and ask, by examining their \DETS, if they experience any observable marginalization. We adopt the definition of "marginalized group" as "groups who have not been as successful as others at achieving educational success, students who find their current curriculum either too challenging or not sufficiently demanding". 

\begin{figure}[tb]
  \includegraphics[width=0.42\textwidth]{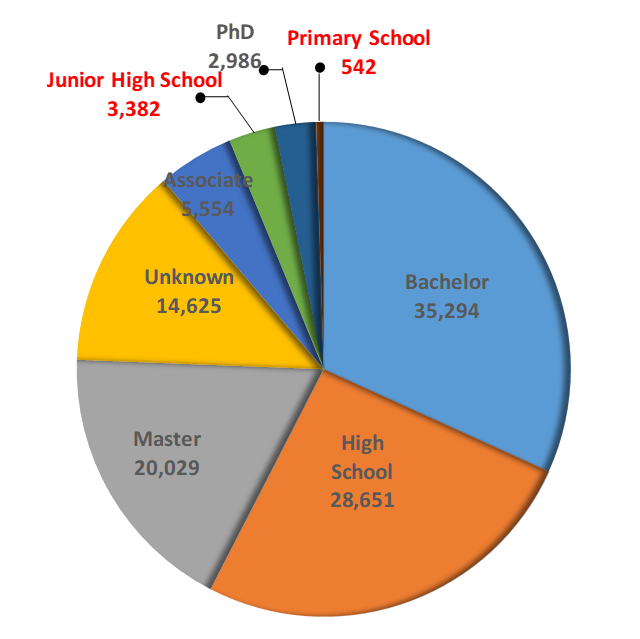}
  \caption{Under-represented education background group in Course \shortcoursename{1A}. Red marks the lower degrees of education.}
  \label{fig:bg_distr_ed}
\end{figure}

We first extract some under-represented groups based on the distribution of highest education level (Figure \ref{fig:bg_distr_ed}) and Gross National Income (GNI) per capita (USD) of country of origin (Table \ref{tab:size}). 

\begin{table}[tb]
  \footnotesize
  \caption{Size of some Potentially Marginalized Student Groups in MITx 6.001, spring 2016}
  \label{tab:size}
  \begin{tabular}{ccc}
    \toprule
    Group Category&Proportion&Absolute Amount\\
    \midrule
    Students with low education degree\footnote{Contains students that declare primary school or junior high school as their highest level of education} &3.5\%&3,924\\
    Students from low-income Economies & 1.4\% &1,547\\
  \bottomrule
\end{tabular}
\end{table}

We use \DETS in the following ways: First, a student's \DETS potentially contains a dropout timestamp\footnote{Dropout time-stamp are e.g. the last time the student access videos, or the last time the student access any materials}. It also contains information about "video replay" behavior, which has one possible interpretation of one "finding certain videos difficult to digest", and thus could indicate where the students need help. Finally, \DET can be intuitive to interpret, so in small quantities we can visually inspect the \DET's of a group. The underrepresented groups we study here are:
\begin{itemize}
\item students whose declared highest education level is either primary school or junior high school
\item students from low-income economy entities classified by World Bank based on GNI per capita \cite{world2016world}
\end{itemize}

\subsection{Students with Low Education Background}
Before MIT released \textit{Introduction to Computational Thinking and Data Science} on Edx, the course was offered in traditional classroom to MIT students, with the prerequisite of the course \textit{Introduction to Computer Science and Programming Using Python} or have the same level of knowledge background. The MOOC course attracts students from all educational backgrounds, with 3.5\% of the students have only primary school or junior high school degrees. Therefore, one concern is whether those students find the course materials too challenging, and, if they do, which part is most challenging. We leverage the latent cognitive meaning of "replaying videos" --- a student plays a lecture video for a second time since the student find the content challenging. We group students considered low in their education background, and count the total number of their replays for each video. As a control, we do the same to the group of student that have college-level education, who account for the majority of the course population (around 60\%). 

\begin{figure}[t]
\includegraphics[width=0.49\textwidth]{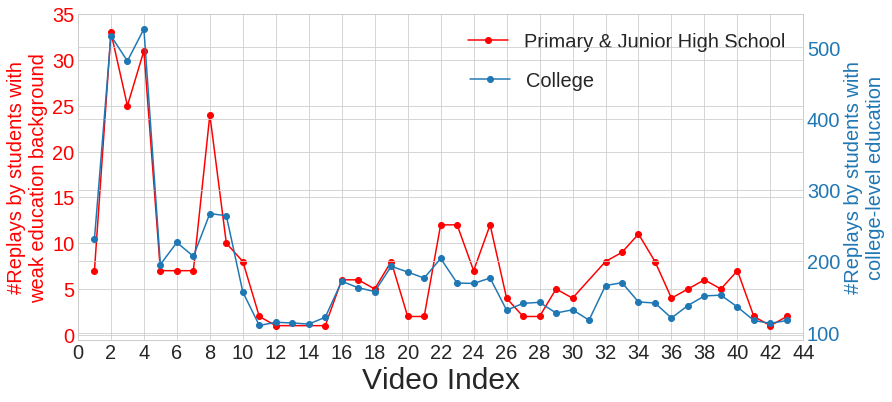}
\caption{The left(red) y-axis shows the target group and the right(blue) shows the majority (baseline). The number of replays follow the same trend. Students with low education background are more sensitive to difficulty changes in videos.}
\label{fig:weak_edu}
\end{figure}

Figure \ref{fig:weak_edu} visualizes the result after scaling y-axis for both groups. We notice that both groups of students follow similar patterns of replaying lecture videos throughout the course. However, the red(low student) line fluctuates more than the blue (control) line does, which indicates that students with low education background at the least are distinctive and could be sensitive to the varying difficulty of course materials. One-tailed t-test is performed on mean absolute values of normalized frequency fluctuations of low-education (target) group and high-education (control) group replay the lecture videos more frequently. We conclude with 99\% confidence level that the target group is more sensitive to difficulty changes in lecture videos.

\begin{figure}[t]
\includegraphics[width=0.35\textwidth]{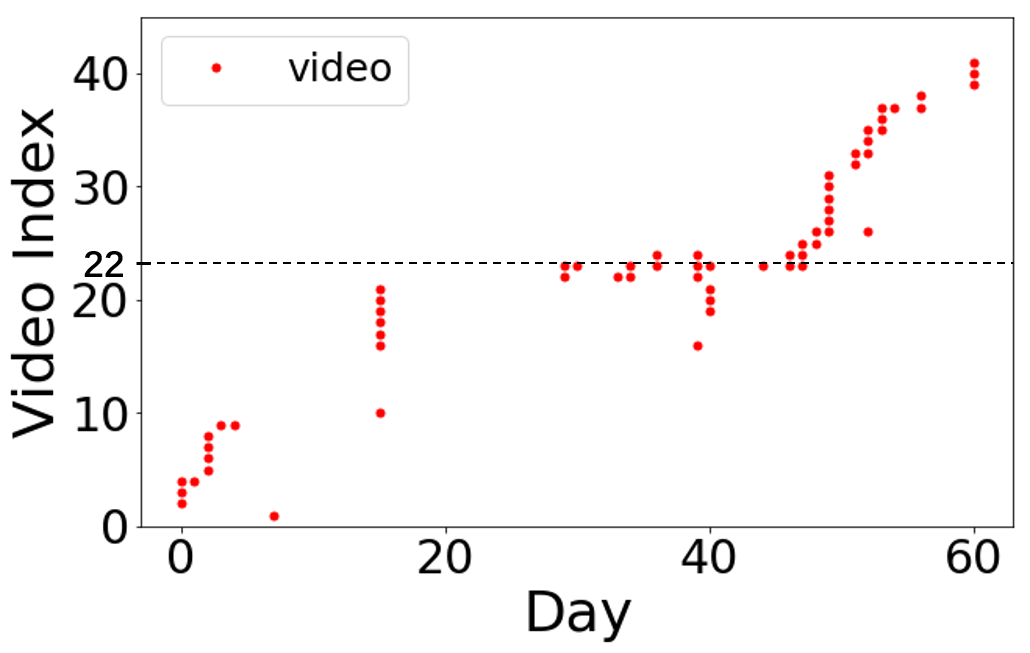}
\caption{\DET of a typical student struggling at around video 22. His highest education degree is junior high school.}
\label{fig:v22}
\end{figure}

This prompts us to investigate videos associated with prominent fluctuations. Video 8 introduces the students to the implementation of graph models and video 22 introduces confidence intervals. Both cover advanced topics that would be rarely touched by students before college, so it makes sense that students with lower education background replay these more. For example, figure \ref{fig:v22} shows a typical student struggling at video 22. The analysis indicates that to help these group of students Edx could provide more reference materials for these two videos. 

\subsection{Students from Low-income Economies}
\begin{figure}[t]
\includegraphics[width=0.49\textwidth]{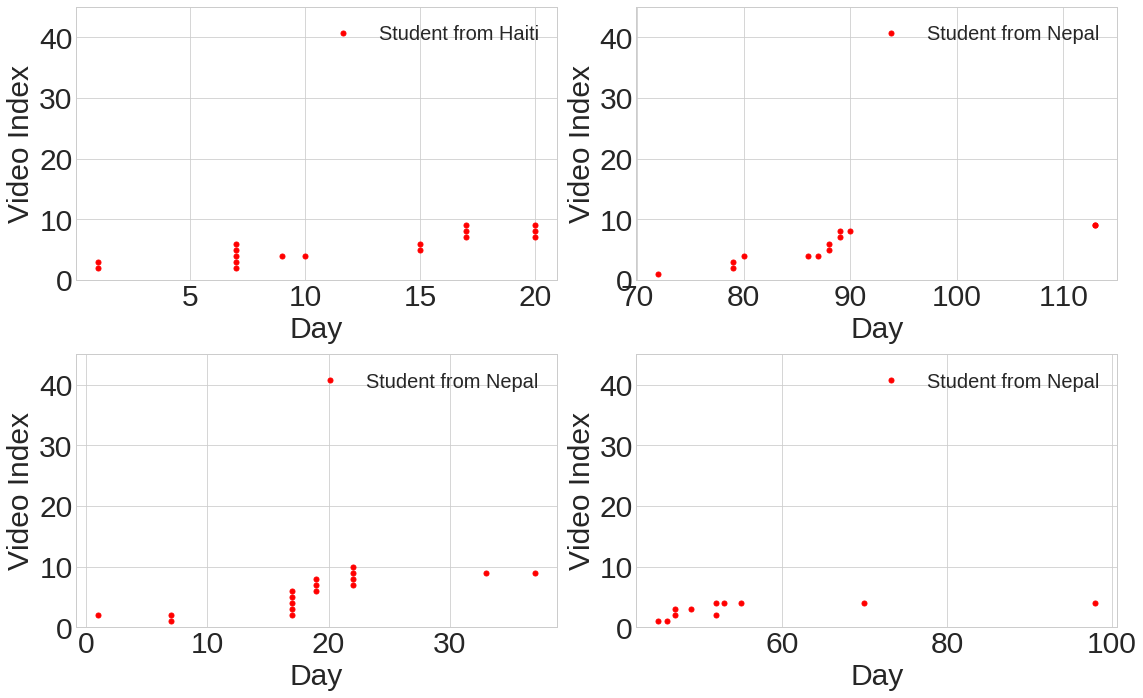}
\caption{\DET of the 4 students that watched the most videos from low-income economies. None of the 182 students from low-income economies managed to proceed beyond half of the lecture videos.}
\label{fig:low_income}
\end{figure}

According to the standard of\cite{world2016world} issued by the World Bank in 2016, students from low-income economies have limited access to education. The lack of computers, adequate bandwidth, and quality educations all pose great challenges for them to perform well in a MOOC. In the largest offering Course \shortcoursename{1A}, We discover that out of the 182 students from low-income economies (Top 3: Uganda, Nepal, Ethiopia), only 50 watched videos at all, and only 4 watched more than 1/4 lecture videos. The four students' video-watching \DET's are shown in Figure \ref{fig:low_income}. All struggle with their learning according to the \DET, even though none dropped out early, nobody proceeded beyond video 11. As a comparison, around 45\% of the average MOOC users in the same offering proceed beyond video 11. This can imply that MOOCs are a form of remote education still inaccessible to many marginalized students and many of our previous dropout analysis fall short of representing some students that are really in need of help.

\section{Student Learning Representation with \DET}
\label{sec:vdag}\label{subsec:2dlearn}

In this section, we present the empirical studies we conduct with \DET that learn 10-dimensional vector representations and visualize data of many students. To examine the effectiveness of our learned representations, we project them to 2D space and visualize them together with the student's grade (encoded by the color). 

\subsection{Traditional Feature Embedding as Baseline}
\label{subsec:scheme1}
\DET supports traditional scheme of features extraction. For video-watching \DET, We devised 10 features to summarize user's video watching behaviors: \textit{n\_unique\_videos} (number of unique videos watched), \textit{n\_days} (number of active days (of video access)), \textit{ave\_day\_intervals} (average length of gaps between two consecutive active days),  \textit{var\_days} (variance of active days), \textit{ave\_video\_intervals} (average length of index gaps between two consecutive watched videos), \textit{var\_videos} (variance of indices of watched videos), \textit{rate\_videos\_repeats} (proportion of replayed videos),\textit{ n\_videos\_per\_day } (average number of videos watched per active day), \textit{var\_day\_intervals} (variance of lengths between two consecutive active days), \textit{var\_video\_intervals} (variance of indices gaps between two consecutive watched videos)

\begin{figure*}[t]
\includegraphics[width=1\textwidth]{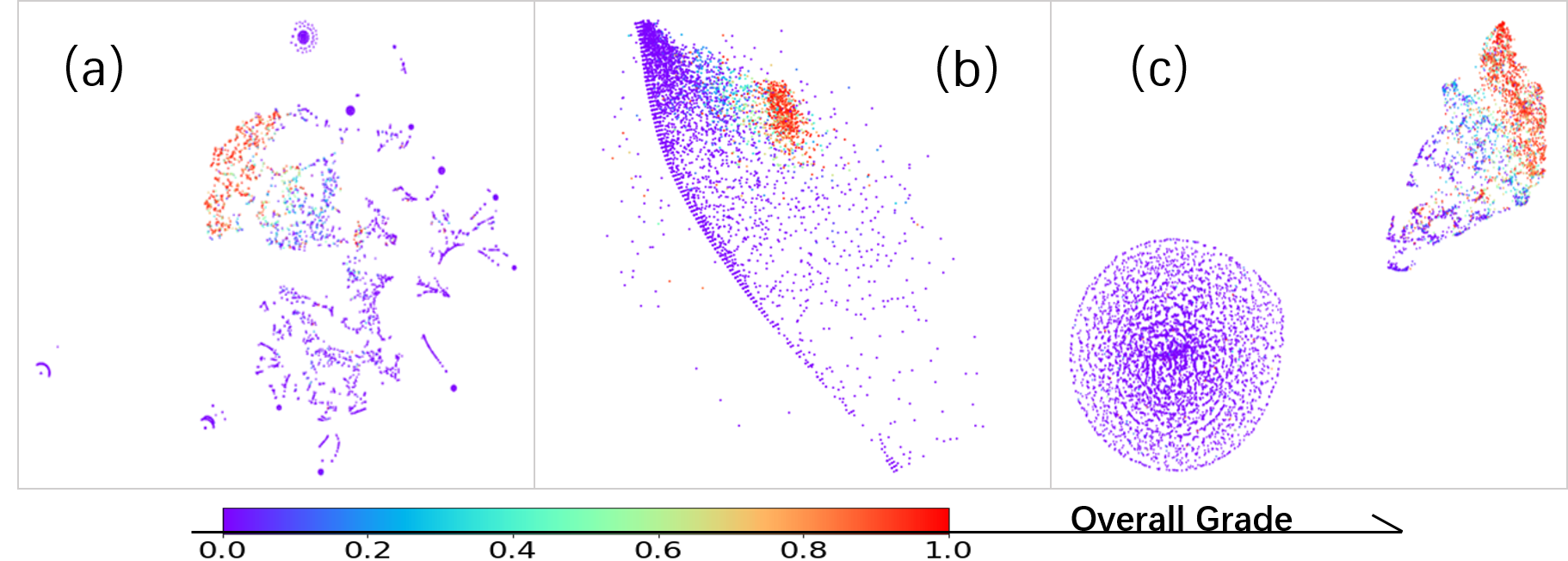}
\caption{(a) Handcrafted feature 2D projection; (b) DTW of \DET 2D projection; (c) CNN-AE of \DET 2D projection. T-SNE 2D representations learned from feature vectors are visualized with corresponding grades encoded by color for all students of MITx 6.00.2x (Course \shortcoursename{2C}).}
\label{fig:clustering}
\end{figure*}

We obtain 10-dimensional feature vectors for each student by calculating corresponding features. The features are standardized to eliminate the bias introduced by the absolute values. To examine the obtained representations, we project the features to 2D spaces with T-SNE and visualize the coordinate vectors with grades encoded by color. The result is shown in figure\ref{fig:clustering}. 

\subsection{Feature Embedding from a Distance Matrix}

The time-series described by \DET can be automatically transformed into features by machine learning. Motivated by \cite{NIPS2013_5021}, we notice that  directly declaring a pool of "good features" to summarize learning behaviors is difficult and sometimes controversial. So, we convert the problem to the challenge of finding a measure of the distance between a pair of two-dimensional time series. Dynamic Time Warping(DTW) proposed by \cite{berndt1994using} serves this purpose well, because it measures the distance between most similar segments in 2D time series. The overall distance is obtained by summing up the pairwise distance between points on most similar segments in the time series, see Figure~\ref{fig:dtw_illustration}. We also normalize the distance by number of matched pairs to eliminate the effect of the time series length. When we perform the distance calculation for each pair of students, a distance matrix of size $n \times n$ is obtained, where $n$ is the total number of students in the course iteration.

Next, with the distance matrix, we use multi-dimension scaling(MDS) to find a set of embeddings (vectors) in 10-dimensional space that best preserve the pairwise distances given by the distance matrix. In this way, the embeddings become representations for a student's learning trajectory.  Again, we project the representations to 2D spaces with T-SNE and visualize the 2D coordinates (vectors), encoding students' grades with colors. Figure \ref{fig:clustering} shows that video-day graphs correlate with a student's grade, with high-grade students concentrating in a cluster. Meanwhile, students with higher grades watch videos in more similar manners than students with lower grades do, leading to the high-grade samples clustering densely at one place while low grade students are scattered. 

More importantly, since DTW considers both content temporal domain information, it takes into consideration more comprehensive information than traditional method does. While studies like \cite{halawa2014dropout} report "recognizing four common persistence patterns that represent the majority of MOOC students", it is also important to point out that technically the inter-cluster distances are not large enough compared to the within-cluster ones. In other words, those prototypes are not observed from the video watching visualization. 

\subsection{Feature Embedding with a Convolutional Autoencoder}
A drawback with DTW is that it essentially "smoothes" all items in the time series, rendering it less sensitive to local video watching patterns (i.e. replaying the same videos a number of days in a row,  watching many videos in a single day, etc.). We are inspired to harness the power of convolutional neural network(CNN) to recognize the local video watching patterns.  We construct a CNN autoencoder(CNN-AE), and treat each student's video-day activity graph as input to the autoencoder. In other words, every two-tuple $(d_i, v_i)$ in the time series maps to a pixel valued 1 at position  $(d_i, v_i)$ on the image. Tables \ref{tab:cnn_ae} in Appendix gives the CNN-AE's structure. 

\begin{figure}[t]
\subfloat[Original \DET(left), CNN-AE reconstruction(Right). Student with a long trajectory]{\includegraphics[width=0.48\textwidth]{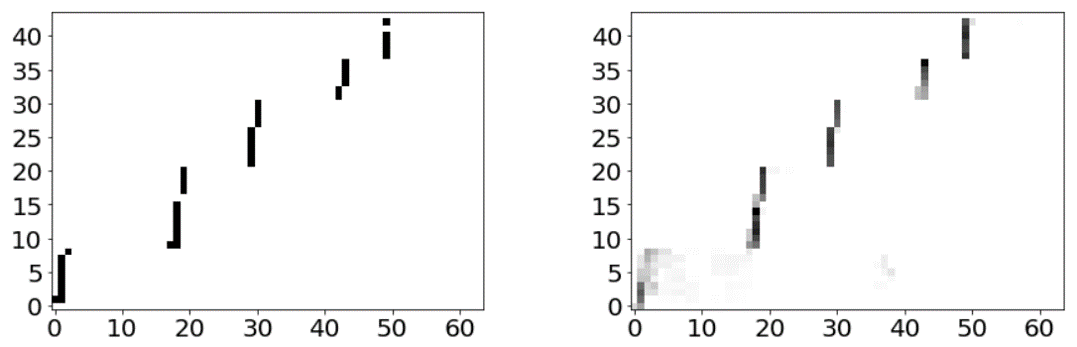}\label{fig:reconstruct1}}\\
\subfloat[Original \DET(left), CNN-AE reconstruction(Right). Student with a shorter trajectory]{\includegraphics[width=0.48\textwidth]{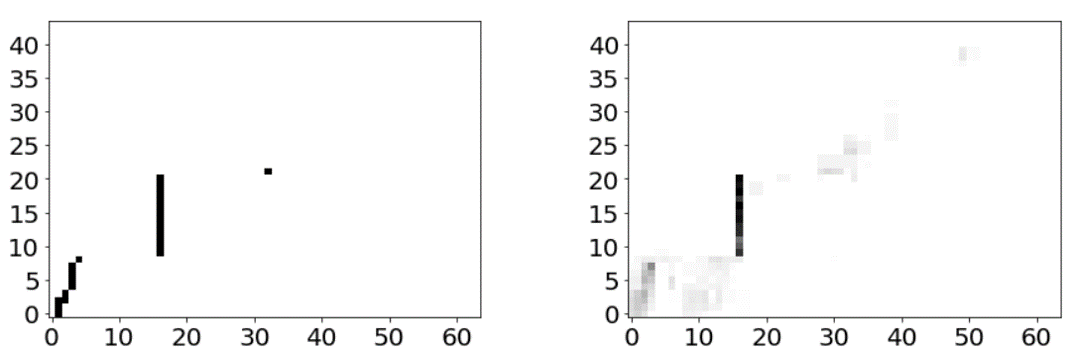}\label{fig:reconstruct4}}
\caption{Original vs. CNN-AE-reconstructed video-day examples. Original is to the left and reconstructed to the right. The CNN-AE reconstruction captures the major trajectory.}
\label{fig:reconstruct}
\end{figure}

We encode the activity to 10 dimensions and then decode it with the aim to condense the most important information in the reduced embedding. The embedded vector representation is then treated the same as the feature vector obtained in Section~\ref{subsec:scheme1} --- being projected to 2D space and plotted. Figure \ref{fig:clustering} plots the 2D embeddings encoded by course grade on unit scale, with high grade students forming one cluster. Figure \ref{fig:reconstruct} shows two typical video-day graphs and their reconstructed counterpart. 

Figure \ref{fig:reconstruct1} shows that the CNN-AE is capable of reconstructing the closest video watching patterns of the video-day graphs. Despite visible blurring and missing pixels, the reconstruction result is by and large satisfactory. The CNN-AE do not form multiple clusters. It is also noticed from (a) and (c) of  the figure that projections form clear separation at around grade = 0.6, which is the passing line of the course. People's learning behaviors become more distinct when they approach the passing grade. Some of them give up and their projections go to one side, while others persist through and their projections move to the other side.

\section{Conclusions \& Future Work}
\label{sec:conclusions--future}

In this paper, we introduce a new perspective of learning behavior analytics in MOOC's: the detailed access trajectory. Its underlying two-dimensional time series facilitate quantitative analysis of learning behaviors. We demonstrate \DET's research value via a empirical studies including representation on two courses regarding introduction to programming and computational thinking. 

We observe Learning Behavior Patterns from video watching with \DETS and observe three distinctive behavior patterns that are correlated with high grade: 
\begin{inparaenum}
\item \reviewAfterALongTime,
\item \returnToPreviouslyWatched,
\item \returnToPreviouslySkipped.
\end{inparaenum}
\DETS are also capable of informing learning designers about LDPs, a core concern of designers and instructors. We find evidence of an LDP regarding students watching a video and then going to the forum, but no significant correlation with grade is found for students that exhibits this behavior. In addition, we use \DETS to examine students from educational and geographical backgrounds that make them potentially marginalized. The \DET helps identify specific videos students with lower educational background are struggling with, as well as observing that students watching MOOC videos from countries with low GNI watch very few videos. Finally, we explored summarizing \DETS and mapping them into a low dimensional embedding for visualization and clustering.

In the future, a more systematic examination combinations of the three categories of \DET can be performed. More LDPs can be identified and queried. In addition, dropout labeling is enabled by \DET, and an interesting topic going forward is to look for local learning patterns that indicate the dropout in the near future. Finally, there is more work to be done on under-represented groups of students.

\bibliographystyle{ACM-Reference-Format}
\bibliography{references}


\begin{thebibliography}{29}


\ifx \showCODEN    \undefined \def \showCODEN     #1{\unskip}     \fi
\ifx \showDOI      \undefined \def \showDOI       #1{#1}\fi
\ifx \showISBNx    \undefined \def \showISBNx     #1{\unskip}     \fi
\ifx \showISBNxiii \undefined \def \showISBNxiii  #1{\unskip}     \fi
\ifx \showISSN     \undefined \def \showISSN      #1{\unskip}     \fi
\ifx \showLCCN     \undefined \def \showLCCN      #1{\unskip}     \fi
\ifx \shownote     \undefined \def \shownote      #1{#1}          \fi
\ifx \showarticletitle \undefined \def \showarticletitle #1{#1}   \fi
\ifx \showURL      \undefined \def \showURL       {\relax}        \fi
\providecommand\bibfield[2]{#2}
\providecommand\bibinfo[2]{#2}
\providecommand\natexlab[1]{#1}
\providecommand\showeprint[2][]{arXiv:#2}

\bibitem[\protect\citeauthoryear{Alexander}{Alexander}{1979}]%
        {alexander1979timeless}
\bibfield{author}{\bibinfo{person}{Christopher Alexander}.}
  \bibinfo{year}{1979}\natexlab{}.
\newblock \bibinfo{booktitle}{\emph{The timeless way of building}}.
  Vol.~\bibinfo{volume}{1}.
\newblock \bibinfo{publisher}{Oxford University Press}, \bibinfo{address}{New
  York, New York}.
\newblock


\bibitem[\protect\citeauthoryear{Berndt and Clifford}{Berndt and
  Clifford}{1994}]%
        {berndt1994using}
\bibfield{author}{\bibinfo{person}{Donald~J Berndt} {and}
  \bibinfo{person}{James Clifford}.} \bibinfo{year}{1994}\natexlab{}.
\newblock \showarticletitle{Using dynamic time warping to find patterns in time
  series.}. In \bibinfo{booktitle}{\emph{KDD workshop}}.
  \bibinfo{publisher}{AAAI Press}, \bibinfo{address}{44 West 4th Street, New
  York, New York 10012-1126}, \bibinfo{pages}{359--370}.
\newblock


\bibitem[\protect\citeauthoryear{Boroujeni and Dillenbourg}{Boroujeni and
  Dillenbourg}{2018}]%
        {boroujeni2018discovery}
\bibfield{author}{\bibinfo{person}{Mina~Shirvani Boroujeni} {and}
  \bibinfo{person}{Pierre Dillenbourg}.} \bibinfo{year}{2018}\natexlab{}.
\newblock \showarticletitle{Discovery and temporal analysis of latent study
  patterns in MOOC interaction sequences}. In
  \bibinfo{booktitle}{\emph{Proceedings of the 8th International Conference on
  Learning Analytics and Knowledge}} \emph{(\bibinfo{series}{LAK '18})}.
  \bibinfo{publisher}{ACM}, \bibinfo{address}{New York, NY, USA},
  \bibinfo{pages}{206--215}.
\newblock


\bibitem[\protect\citeauthoryear{Boyer and Veeramachaneni}{Boyer and
  Veeramachaneni}{2015}]%
        {10.1007/978-3-319-19773-9_6}
\bibfield{author}{\bibinfo{person}{Sebastien Boyer} {and}
  \bibinfo{person}{Kalyan Veeramachaneni}.} \bibinfo{year}{2015}\natexlab{}.
\newblock \showarticletitle{Transfer Learning for Predictive Models in Massive
  Open Online Courses}. In \bibinfo{booktitle}{\emph{Artificial Intelligence in
  Education}}, \bibfield{editor}{\bibinfo{person}{Cristina Conati},
  \bibinfo{person}{Neil Heffernan}, \bibinfo{person}{Antonija Mitrovic}, {and}
  \bibinfo{person}{M.~Felisa Verdejo}} (Eds.). \bibinfo{publisher}{Springer
  International Publishing}, \bibinfo{address}{Cham}, \bibinfo{pages}{54--63}.
\newblock


\bibitem[\protect\citeauthoryear{Brugha and Restoule}{Brugha and
  Restoule}{2016}]%
        {brugha2016examining}
\bibfield{author}{\bibinfo{person}{Meaghan Brugha} {and}
  \bibinfo{person}{Jean-Paul Restoule}.} \bibinfo{year}{2016}\natexlab{}.
\newblock \showarticletitle{Examining the learning networks of a MOOC}.
\newblock \bibinfo{journal}{\emph{Data Mining and Learning Analytics:
  Applications in Educational Research}} (\bibinfo{year}{2016}),
  \bibinfo{pages}{121}.
\newblock


\bibitem[\protect\citeauthoryear{Corrin, de~Barba, and Bakharia}{Corrin
  et~al\mbox{.}}{2017}]%
        {Corrin:2017:ULA:3027385.3027448}
\bibfield{author}{\bibinfo{person}{Linda Corrin}, \bibinfo{person}{Paula~G. de
  Barba}, {and} \bibinfo{person}{Aneesha Bakharia}.}
  \bibinfo{year}{2017}\natexlab{}.
\newblock \showarticletitle{Using Learning Analytics to Explore Help-seeking
  Learner Profiles in MOOCs}. In \bibinfo{booktitle}{\emph{Proceedings of the
  Seventh International Learning Analytics \&\#38; Knowledge Conference}}
  \emph{(\bibinfo{series}{LAK '17})}. \bibinfo{publisher}{ACM},
  \bibinfo{address}{New York, NY, USA}, \bibinfo{pages}{424--428}.
\newblock
\showISBNx{978-1-4503-4870-6}
\urldef\tempurl%
\url{https://doi.org/10.1145/3027385.3027448}
\showDOI{\tempurl}


\bibitem[\protect\citeauthoryear{Davis, Kizilcec, Hauff, and Houben}{Davis
  et~al\mbox{.}}{2018}]%
        {Davis:2018:HMK:3170358.3170383}
\bibfield{author}{\bibinfo{person}{Dan Davis}, \bibinfo{person}{Ren{\'e}~F.
  Kizilcec}, \bibinfo{person}{Claudia Hauff}, {and} \bibinfo{person}{Geert-Jan
  Houben}.} \bibinfo{year}{2018}\natexlab{}.
\newblock \showarticletitle{The Half-life of MOOC Knowledge: A Randomized Trial
  Evaluating Knowledge Retention and Retrieval Practice in MOOCs}. In
  \bibinfo{booktitle}{\emph{Proceedings of the 8th International Conference on
  Learning Analytics and Knowledge}} \emph{(\bibinfo{series}{LAK '18})}.
  \bibinfo{publisher}{ACM}, \bibinfo{address}{New York, NY, USA},
  \bibinfo{pages}{1--10}.
\newblock
\showISBNx{978-1-4503-6400-3}
\urldef\tempurl%
\url{https://doi.org/10.1145/3170358.3170383}
\showDOI{\tempurl}


\bibitem[\protect\citeauthoryear{DeBoer, Stump, Seaton, and Breslow}{DeBoer
  et~al\mbox{.}}{2013a}]%
        {deboer2013diversity}
\bibfield{author}{\bibinfo{person}{Jennifer DeBoer}, \bibinfo{person}{Glenda~S
  Stump}, \bibinfo{person}{Daniel Seaton}, {and} \bibinfo{person}{Lori
  Breslow}.} \bibinfo{year}{2013}\natexlab{a}.
\newblock \showarticletitle{Diversity in MOOC students’ backgrounds and
  behaviors in relationship to performance in 6.002 x}. In
  \bibinfo{booktitle}{\emph{Proceedings of the Sixth Learning International
  Networks Consortium Conference}}, Vol.~\bibinfo{volume}{4}.
  \bibinfo{pages}{16--19}.
\newblock


\bibitem[\protect\citeauthoryear{DeBoer, Stump, Seaton, Ho, Pritchard, and
  Breslow}{DeBoer et~al\mbox{.}}{2013b}]%
        {deboer2013bringing}
\bibfield{author}{\bibinfo{person}{Jennifer DeBoer}, \bibinfo{person}{Glenda~S
  Stump}, \bibinfo{person}{Daniel Seaton}, \bibinfo{person}{Andrew Ho},
  \bibinfo{person}{David~E Pritchard}, {and} \bibinfo{person}{Lori Breslow}.}
  \bibinfo{year}{2013}\natexlab{b}.
\newblock \showarticletitle{Bringing student backgrounds online: MOOC user
  demographics, site usage, and online learning}. In
  \bibinfo{booktitle}{\emph{Educational Data Mining 2013}}.
\newblock


\bibitem[\protect\citeauthoryear{Elbadrawy, Polyzou, Ren, Sweeney, Karypis, and
  Rangwala}{Elbadrawy et~al\mbox{.}}{2016}]%
        {7452320}
\bibfield{author}{\bibinfo{person}{A. Elbadrawy}, \bibinfo{person}{A. Polyzou},
  \bibinfo{person}{Z. Ren}, \bibinfo{person}{M. Sweeney}, \bibinfo{person}{G.
  Karypis}, {and} \bibinfo{person}{H. Rangwala}.}
  \bibinfo{year}{2016}\natexlab{}.
\newblock \showarticletitle{Predicting Student Performance Using Personalized
  Analytics}.
\newblock \bibinfo{journal}{\emph{Computer}} \bibinfo{volume}{49},
  \bibinfo{number}{4} (\bibinfo{date}{Apr} \bibinfo{year}{2016}),
  \bibinfo{pages}{61--69}.
\newblock
\showISSN{0018-9162}
\urldef\tempurl%
\url{https://doi.org/10.1109/MC.2016.119}
\showDOI{\tempurl}


\bibitem[\protect\citeauthoryear{Ferguson and Clow}{Ferguson and Clow}{2015}]%
        {Ferguson:2015:EEA:2723576.2723606}
\bibfield{author}{\bibinfo{person}{Rebecca Ferguson} {and}
  \bibinfo{person}{Doug Clow}.} \bibinfo{year}{2015}\natexlab{}.
\newblock \showarticletitle{Examining Engagement: Analysing Learner
  Subpopulations in Massive Open Online Courses (MOOCs)}. In
  \bibinfo{booktitle}{\emph{Proceedings of the Fifth International Conference
  on Learning Analytics And Knowledge}} \emph{(\bibinfo{series}{LAK '15})}.
  \bibinfo{publisher}{ACM}, \bibinfo{address}{New York, NY, USA},
  \bibinfo{pages}{51--58}.
\newblock
\showISBNx{978-1-4503-3417-4}
\urldef\tempurl%
\url{https://doi.org/10.1145/2723576.2723606}
\showDOI{\tempurl}


\bibitem[\protect\citeauthoryear{Group}{Group}{2016}]%
        {world2016world}
\bibfield{author}{\bibinfo{person}{World~Bank Group}.}
  \bibinfo{year}{2016}\natexlab{}.
\newblock \bibinfo{booktitle}{\emph{World development indicators 2016}}.
\newblock \bibinfo{publisher}{World Bank Publications}.
\newblock


\bibitem[\protect\citeauthoryear{Guo and Reinecke}{Guo and Reinecke}{2014}]%
        {guo2014demographic}
\bibfield{author}{\bibinfo{person}{Philip~J Guo} {and}
  \bibinfo{person}{Katharina Reinecke}.} \bibinfo{year}{2014}\natexlab{}.
\newblock \showarticletitle{Demographic differences in how students navigate
  through MOOCs}. In \bibinfo{booktitle}{\emph{Proceedings of the first ACM
  conference on Learning@ scale conference}}. ACM, \bibinfo{pages}{21--30}.
\newblock


\bibitem[\protect\citeauthoryear{Halawa, Greene, and Mitchell}{Halawa
  et~al\mbox{.}}{2014}]%
        {halawa2014dropout}
\bibfield{author}{\bibinfo{person}{Sherif Halawa}, \bibinfo{person}{Daniel
  Greene}, {and} \bibinfo{person}{John Mitchell}.}
  \bibinfo{year}{2014}\natexlab{}.
\newblock \showarticletitle{Dropout prediction in MOOCs using learner activity
  features}.
\newblock \bibinfo{journal}{\emph{Proceedings of the Second European MOOC
  Stakeholder Summit}} (\bibinfo{year}{2014}), \bibinfo{pages}{58--65}.
\newblock


\bibitem[\protect\citeauthoryear{He, Bailey, Rubinstein, and Zhang}{He
  et~al\mbox{.}}{2015}]%
        {He:2015:IAS:2886521.2886563}
\bibfield{author}{\bibinfo{person}{Jiazhen He}, \bibinfo{person}{James Bailey},
  \bibinfo{person}{Benjamin I.~P. Rubinstein}, {and} \bibinfo{person}{Rui
  Zhang}.} \bibinfo{year}{2015}\natexlab{}.
\newblock \showarticletitle{Identifying At-risk Students in Massive Open Online
  Courses}. In \bibinfo{booktitle}{\emph{Proceedings of the Twenty-Ninth AAAI
  Conference on Artificial Intelligence}} \emph{(\bibinfo{series}{AAAI'15})}.
  \bibinfo{publisher}{AAAI Press}, \bibinfo{pages}{1749--1755}.
\newblock
\showISBNx{0-262-51129-0}
\urldef\tempurl%
\url{http://dl.acm.org/citation.cfm?id=2886521.2886563}
\showURL{%
\tempurl}


\bibitem[\protect\citeauthoryear{Hood, Littlejohn, and Milligan}{Hood
  et~al\mbox{.}}{2015}]%
        {HOOD201583}
\bibfield{author}{\bibinfo{person}{Nina Hood}, \bibinfo{person}{Allison
  Littlejohn}, {and} \bibinfo{person}{Colin Milligan}.}
  \bibinfo{year}{2015}\natexlab{}.
\newblock \showarticletitle{Context counts: How learners' contexts influence
  learning in a MOOC}.
\newblock \bibinfo{journal}{\emph{Computers \& Education}}
  \bibinfo{volume}{91} (\bibinfo{year}{2015}), \bibinfo{pages}{83 -- 91}.
\newblock
\showISSN{0360-1315}
\urldef\tempurl%
\url{https://doi.org/10.1016/j.compedu.2015.10.019}
\showDOI{\tempurl}


\bibitem[\protect\citeauthoryear{Kizilcec, Piech, and Schneider}{Kizilcec
  et~al\mbox{.}}{2013}]%
        {Kizilcec:2013:DDA:2460296.2460330}
\bibfield{author}{\bibinfo{person}{Ren{\'e}~F. Kizilcec},
  \bibinfo{person}{Chris Piech}, {and} \bibinfo{person}{Emily Schneider}.}
  \bibinfo{year}{2013}\natexlab{}.
\newblock \showarticletitle{Deconstructing Disengagement: Analyzing Learner
  Subpopulations in Massive Open Online Courses}. In
  \bibinfo{booktitle}{\emph{Proceedings of the Third International Conference
  on Learning Analytics and Knowledge}} \emph{(\bibinfo{series}{LAK '13})}.
  \bibinfo{publisher}{ACM}, \bibinfo{address}{New York, NY, USA},
  \bibinfo{pages}{170--179}.
\newblock
\showISBNx{978-1-4503-1785-6}
\urldef\tempurl%
\url{https://doi.org/10.1145/2460296.2460330}
\showDOI{\tempurl}


\bibitem[\protect\citeauthoryear{Laurillard}{Laurillard}{2013}]%
        {laurillard2013teaching}
\bibfield{author}{\bibinfo{person}{Diana Laurillard}.}
  \bibinfo{year}{2013}\natexlab{}.
\newblock \bibinfo{booktitle}{\emph{Teaching as a design science: Building
  pedagogical patterns for learning and technology}}.
\newblock \bibinfo{publisher}{Routledge}.
\newblock


\bibitem[\protect\citeauthoryear{Law, Li, Herrera, Chan, and Pong}{Law
  et~al\mbox{.}}{2017}]%
        {law2017pattern}
\bibfield{author}{\bibinfo{person}{Nancy Law}, \bibinfo{person}{Ling Li},
  \bibinfo{person}{Liliana~Farias Herrera}, \bibinfo{person}{Andy Chan}, {and}
  \bibinfo{person}{Ting-Chuen Pong}.} \bibinfo{year}{2017}\natexlab{}.
\newblock \showarticletitle{A Pattern Language Based Learning Design Studio for
  an Analytics Informed Inter-Professional Design Community}.
\newblock \bibinfo{journal}{\emph{Interaction Design and Architecture (s)}}
  (\bibinfo{year}{2017}), \bibinfo{pages}{92}.
\newblock


\bibitem[\protect\citeauthoryear{McAndrew and Scanlon}{McAndrew and
  Scanlon}{2013}]%
        {mcandrew2013open}
\bibfield{author}{\bibinfo{person}{Patrick McAndrew} {and}
  \bibinfo{person}{Eileen Scanlon}.} \bibinfo{year}{2013}\natexlab{}.
\newblock \showarticletitle{Open learning at a distance: lessons for struggling
  MOOCs}.
\newblock \bibinfo{journal}{\emph{Science}} \bibinfo{volume}{342},
  \bibinfo{number}{6165} (\bibinfo{year}{2013}), \bibinfo{pages}{1450--1451}.
\newblock


\bibitem[\protect\citeauthoryear{Meier, Xu, Atan, and van~der Schaar}{Meier
  et~al\mbox{.}}{2016}]%
        {7313031}
\bibfield{author}{\bibinfo{person}{Y. Meier}, \bibinfo{person}{J. Xu},
  \bibinfo{person}{O. Atan}, {and} \bibinfo{person}{M. van~der Schaar}.}
  \bibinfo{year}{2016}\natexlab{}.
\newblock \showarticletitle{Predicting Grades}.
\newblock \bibinfo{journal}{\emph{IEEE Transactions on Signal Processing}}
  \bibinfo{volume}{64}, \bibinfo{number}{4} (\bibinfo{date}{Feb}
  \bibinfo{year}{2016}), \bibinfo{pages}{959--972}.
\newblock
\showISSN{1053-587X}
\urldef\tempurl%
\url{https://doi.org/10.1109/TSP.2015.2496278}
\showDOI{\tempurl}


\bibitem[\protect\citeauthoryear{Mikolov, Sutskever, Chen, Corrado, and
  Dean}{Mikolov et~al\mbox{.}}{2013}]%
        {NIPS2013_5021}
\bibfield{author}{\bibinfo{person}{Tomas Mikolov}, \bibinfo{person}{Ilya
  Sutskever}, \bibinfo{person}{Kai Chen}, \bibinfo{person}{Greg~S Corrado},
  {and} \bibinfo{person}{Jeff Dean}.} \bibinfo{year}{2013}\natexlab{}.
\newblock \showarticletitle{Distributed Representations of Words and Phrases
  and their Compositionality}.
\newblock In \bibinfo{booktitle}{\emph{Advances in Neural Information
  Processing Systems 26}}, \bibfield{editor}{\bibinfo{person}{C.~J.~C. Burges},
  \bibinfo{person}{L.~Bottou}, \bibinfo{person}{M.~Welling},
  \bibinfo{person}{Z.~Ghahramani}, {and} \bibinfo{person}{K.~Q. Weinberger}}
  (Eds.). \bibinfo{publisher}{Curran Associates, Inc.},
  \bibinfo{pages}{3111--3119}.
\newblock
\urldef\tempurl%
\url{http://papers.nips.cc/paper/5021-distributed-representations-of-words-and-phrases-and-their-compositionality.pdf}
\showURL{%
\tempurl}


\bibitem[\protect\citeauthoryear{Pursel, Zhang, Jablokow, Choi, and
  Velegol}{Pursel et~al\mbox{.}}{2016}]%
        {pursel2016understanding}
\bibfield{author}{\bibinfo{person}{Barton~K Pursel}, \bibinfo{person}{L Zhang},
  \bibinfo{person}{Kathryn~W Jablokow}, \bibinfo{person}{GW Choi}, {and}
  \bibinfo{person}{D Velegol}.} \bibinfo{year}{2016}\natexlab{}.
\newblock \showarticletitle{Understanding MOOC students: motivations and
  behaviours indicative of MOOC completion}.
\newblock \bibinfo{journal}{\emph{Journal of Computer Assisted Learning}}
  \bibinfo{volume}{32}, \bibinfo{number}{3} (\bibinfo{year}{2016}),
  \bibinfo{pages}{202--217}.
\newblock


\bibitem[\protect\citeauthoryear{Rai and Chunrao}{Rai and Chunrao}{2016}]%
        {rai2016influencing}
\bibfield{author}{\bibinfo{person}{Laxmisha Rai} {and} \bibinfo{person}{Deng
  Chunrao}.} \bibinfo{year}{2016}\natexlab{}.
\newblock \showarticletitle{Influencing factors of success and failure in MOOC
  and general analysis of learner behavior}.
\newblock \bibinfo{journal}{\emph{International Journal of Information and
  Education Technology}} \bibinfo{volume}{6}, \bibinfo{number}{4}
  (\bibinfo{year}{2016}), \bibinfo{pages}{262}.
\newblock


\bibitem[\protect\citeauthoryear{Ren, Rangwala, and Johri}{Ren
  et~al\mbox{.}}{2016}]%
        {ren2016predict}
\bibfield{author}{\bibinfo{person}{Zhiyun Ren}, \bibinfo{person}{Huzefa
  Rangwala}, {and} \bibinfo{person}{Aditya Johri}.}
  \bibinfo{year}{2016}\natexlab{}.
\newblock \showarticletitle{Predicting Performance on MOOC Assessments using
  Multi-Regression Models}.
\newblock  (\bibinfo{date}{05} \bibinfo{year}{2016}).
\newblock


\bibitem[\protect\citeauthoryear{Wilson}{Wilson}{2018}]%
        {wilson2018online}
\bibfield{author}{\bibinfo{person}{Mary Wilson}.}
  \bibinfo{year}{2018}\natexlab{}.
\newblock \showarticletitle{Online instructional design in the new world:
  Beyond Gagn{\'e}, Briggs and Wager}.
\newblock  (\bibinfo{year}{2018}).
\newblock


\bibitem[\protect\citeauthoryear{Wise and Cui}{Wise and Cui}{2018}]%
        {Wise:2018:URD:3170358.3170403}
\bibfield{author}{\bibinfo{person}{Alyssa~Friend Wise} {and}
  \bibinfo{person}{Yi Cui}.} \bibinfo{year}{2018}\natexlab{}.
\newblock \showarticletitle{Unpacking the Relationship Between Discussion Forum
  Participation and Learning in MOOCs: Content is Key}. In
  \bibinfo{booktitle}{\emph{Proceedings of the 8th International Conference on
  Learning Analytics and Knowledge}} \emph{(\bibinfo{series}{LAK '18})}.
  \bibinfo{publisher}{ACM}, \bibinfo{address}{New York, NY, USA},
  \bibinfo{pages}{330--339}.
\newblock
\showISBNx{978-1-4503-6400-3}
\urldef\tempurl%
\url{https://doi.org/10.1145/3170358.3170403}
\showDOI{\tempurl}


\bibitem[\protect\citeauthoryear{Xu and Yang}{Xu and Yang}{2016}]%
        {Xu:2016:MCG:2934357.2934361}
\bibfield{author}{\bibinfo{person}{Bin Xu} {and} \bibinfo{person}{Dan Yang}.}
  \bibinfo{year}{2016}\natexlab{}.
\newblock \showarticletitle{Motivation Classification and Grade Prediction for
  MOOCs Learners}.
\newblock \bibinfo{journal}{\emph{Intell. Neuroscience}}
  \bibinfo{volume}{2016}, Article \bibinfo{articleno}{4} (\bibinfo{date}{Jan.}
  \bibinfo{year}{2016}), \bibinfo{numpages}{1}~pages.
\newblock
\showISSN{1687-5265}
\urldef\tempurl%
\url{https://doi.org/10.1155/2016/2174613}
\showDOI{\tempurl}


\bibitem[\protect\citeauthoryear{Zhuoxuan, Yan, and Xiaoming}{Zhuoxuan
  et~al\mbox{.}}{2015}]%
        {zhuoxuan2015learning}
\bibfield{author}{\bibinfo{person}{Jiang Zhuoxuan}, \bibinfo{person}{Zhang
  Yan}, {and} \bibinfo{person}{Li Xiaoming}.} \bibinfo{year}{2015}\natexlab{}.
\newblock \showarticletitle{Learning behavior analysis and prediction based on
  MOOC data}.
\newblock \bibinfo{journal}{\emph{Journal of computer research and
  development}} \bibinfo{volume}{52}, \bibinfo{number}{3}
  (\bibinfo{year}{2015}), \bibinfo{pages}{614--28}.
\newblock


\end{thebibliography}

\appendix
\section*{Appendix}

\begin{figure}[tb]
\includegraphics[width=0.39\textwidth]{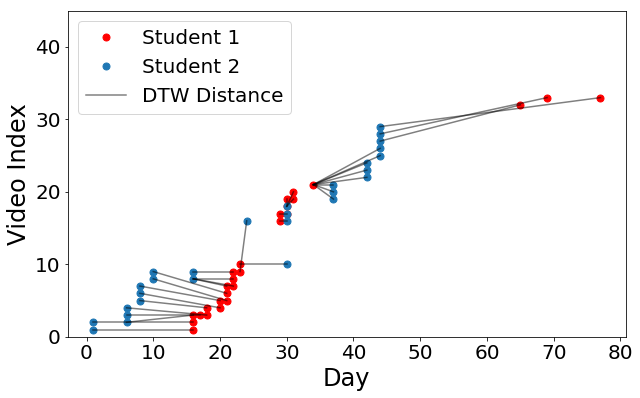}
\caption{Dynamic Time Warping sums up euclidean distances between most similar segments in both time series.}
\label{fig:dtw_illustration}
\end{figure}

\begin{table}[tb]
  \scriptsize
  \caption{Convolutional Autoencoder Structure for Course \shortcoursename{2A}, \shortcoursename{2B}, \shortcoursename{2C}. Activation function: ReLu. Input \DAGs are zero-padded to $44\times 64$.}
  \label{tab:cnn_ae}
  \begin{tabular}{cccc}
    \toprule
 Layer & Size-in & Size-out & Kernel(, Stride) \\ [0.5ex] 
 \midrule
 \midrule
 Conv1 &$44\times 64\times 1$ & $22\times 32\times 16$ & ($2\times 2$), $2$ \\ 
 Conv2 &$22\times 32\times 16$ & $11\times 16\times 32$ & ($2\times 2$), $2$ \\ 
 Conv3 &$11\times 16\times 32$ & $5\times 8\times 64$ & ($2\times 2$), $2$ \\ 
 Conv4 &$5\times 8\times 64$ & $2\times 4\times 128$ & ($2\times 2$), $2$ \\ 
 \midrule
 Fc1 &$1024$ & $10$ & - \\
 Fc2 &$10$ & $1024$ & - \\
  \midrule
 ConvTranspose1 &$2\times 4\times 128$ & $5\times 8\times 64$ & ($3\times 2$), $2$ \\ 
 ConvTranspose2 &$5\times 8\times 64$ & $11\times 16\times 32$ & ($3\times 2$), $2$ \\ 
 ConvTranspose3 &$11\times 16\times 32$ & $22\times 32\times 16$ & ($2\times 2$), $2$ \\ 
 ConvTranspose4 &$22\times 32\times 16$ & $44\times 64\times 1$ & ($2\times 2$), $2$ \\[0.5ex]
  \bottomrule
\end{tabular}
\end{table}

\end{document}